\documentstyle[osa,manuscript]{revtex}

\begin{document}

\title{Non-perturbative results for the spectrum \\
       of surface-disordered waveguides}

\author{N.~M.~Makarov and A.~V.~Moroz}

\address{Institute for Radiophysics and Electronics,
12 Acad.~Proskura St., Kharkov, 310085, Ukraine}

\maketitle

\begin{abstract}
We calculated the spectrum of normal scalar waves in a planar waveguide
with absolutely soft randomly rough boundaries beyond the perturbation
theories in the roughness heights and slopes, basing on the {\it exact}
boundary scattering potential. The spectrum is proved to be a nearly {\it
real non-analytic} function of the dispersion $\zeta^2$ of the roughness
heights (with square-root singularity) as $\zeta^2 \to 0$. The opposite
case of large boundary defects is summarized.
\end{abstract}

In finite systems a long-distance (waveguide) signal propagation is
caused by multiple reflections of the signal from opposite lateral
boundaries. If the boundaries are irregular, each of the reflections is
accompanied by a non-coherent scattering of the travelling wave. The
multiple successive scattering events lead to substantial dephasing and
attenuation of the primary signal.

As far as we know, this effect was first consistently treated in
works~\cite{BFF,BF}. That simple and physically clear approach was based
on the perturbation theory in the squared r.m.s. height $\zeta$ of the
boundary roughness and therefore required small enough $\zeta$, viz.
$(k_z\zeta)^2 \ll 1$, $(\zeta/R_c)^2 \ll 1$, $k\zeta^2/R_c \ll 1$. Here
$k_z$ is the transverse (normal to the waveguiding direction) component of
the wavevector $\vec k$ and $R_c$ is the mean length of the boundary
defects. In more recent papers~\cite{Voron,IPF,DM,Tatarskii} the theory of
wave scattering from statistically rough surfaces was extended to arbitrary
values of the Rayleigh parameter $(k_z\zeta)^2$. However, the other two
inequalities of the above set were still necessary, which made it
impossible to deal with, e.g., steep roughness slopes
($(\zeta/R_c)^2 \gtrsim 1$) and/or the shadowing effect~\cite{BF,Smith}
($k\zeta^2/R_c \gtrsim 1$).

In this Letter we put forward the approach which is {\it non-perturbative}
in the roughness heights and slopes. It is based instead on the
exploitation of the {\it exact} boundary scattering operator. Due to this
fact, we managed to extend the waveguide theory up to the quite general
conditions of weak scattering (\ref{WS}), which are much less restrictive
than the above approximations. The most impressive advantage of our method
is that it leads to new physical results even for the region of small
heights $\zeta$ ($(k\zeta)^2 \ll 1$), where the waveguide propagation is
believed to be well studied. The most surprising result is a {\it
non-analytic} (square-root) dependence of the waveguide spectrum on the
dispersion $\zeta^2$ (i.e. on the coefficient of correlation) of the
roughness heights. This non-analyticity {\it can not} be in principle
derived perturbatively. It means that the exact randomly rough boundary
{\it can not} be reduced to the smooth random-impedance one even for the
arbitrarily small irregularities.

We consider a $2D$ (planar) strip confined to a region of the $xz$-plane
defined by $\xi(x) \leq z \leq d$, where $\xi(x)$ is a Gaussian
distributed random function of the longitudinal coordinate $x$
characterized by the properties

\begin{equation}
\langle\xi(x)\rangle = 0, \qquad
\langle\xi(x)\xi(x^\prime)\rangle = \zeta^2 {\cal W}(|x-x^\prime|).
\label{Gauss}
\end{equation}
The angular brackets denote an average over the ensemble of realizations of
the profile function $\xi(x)$. The binary coefficient of correlation
${\cal W}(|x|)$ has the unit amplitude (${\cal W}(0)=1$) and the typical
width of $R_c$.  The spatial distribution of a scalar wave field inside
the strip is governed by the Helmholtz equation and the temporal
dependence by the factor $\exp(-i\omega t)$, $k=\omega/c$. Both lateral
boundaries $z=\xi(x)$ and $z=d$ are supposed to be absolutely soft, i.e.
the field vanishes upon them. We seek the averaged Green's function
$\langle{\cal G}(x,x^\prime;z,z^\prime)\rangle$ to this Dirichlet boundary
value problem.

The exact integral equation for ${\cal G}(x,x^\prime;z,z^\prime)$ can be
obtained through the use of Green's theorem:

\begin{equation}
{\cal G}(x,x^\prime;z,z^\prime) =
{\cal G}_0(|x-x^\prime|;z,z^\prime)+
\int_{-\infty}^\infty dx_s dz_s
{\cal G}_0(|x-x_s|;z,z_s) \hat\Xi(x_s,z_s)
{\cal G}(x_s,x^\prime;z_s,z^\prime),
\label{GF}
\end{equation}
where $\hat\Xi(x_s,z_s)$ is the {\it effective scattering operator},

\begin{equation}
\hat\Xi(x_s,z_s) = \delta\left[z_s-\xi(x_s)\right]
\left[\frac{\partial}{\partial z_s}-
\frac{d\xi(x_s)}{dx_s}\frac{\partial}{\partial x_s}\right],
\label{Xi}
\end{equation}
and ${\cal G}_0(|x-x^\prime|;z,z^\prime)$ is the Green's function for the
ideal (smooth) strip with $\xi(x) \equiv 0$.

To perform averaging of Eq.~(\ref{GF}) we apply the elegant technique
derived in works~\cite{McGurn,Brown}. Although the averaging procedure is
rather straightforward, it has few fine points which will be discussed
elsewhere~\cite{MM}. As the result, we find the averaged Green's function:

\begin{equation}
\langle{\cal G}(x,x^\prime;z,z^\prime)\rangle \equiv
\overline{\cal G}(|x-x^\prime|;z,z^\prime) =
\int_{-\infty}^\infty\frac{dk_x}{2\pi}
\exp\left[ik_x(x-x^\prime)\right]
\overline G(k_x;z,z^\prime),
\label{G-av}
\end{equation}
where longitudinal Fourier transform $\overline G(k_x;z,z^\prime)$ is given
by

\begin{equation}
\overline G(k_x;z,z^\prime) \simeq
\frac{G_0(k_x;z,z^\prime)}{1-k_z\cot(k_zd)M(k_x)}.
\label{G-Fourier}
\end{equation}
Here $k_z=k_z(k_x)=\sqrt{k^2-k_x^2}$ and $G_0(k_x;z,z^\prime)$ is the
Fourier transform (similar to Eq.~(\ref{G-av})) of
${\cal G}_0(|x-x^\prime|;z,z^\prime)$. The self-energy $M(k_x)$ is defined
via the binary correlator of the scattering operator (\ref{Xi}):

\begin{eqnarray}
M(k_x) & = & \int_{-\infty}^\infty dz_s dz_s^\prime dx_s
\frac{\sin(k_zz_s)}{k_z}\exp(-ik_xx_s)
\nonumber \\
&\times&\langle\hat\Xi(x_s,z_s)
{\cal G}_0(|x_s-x_s^\prime|;z_s,z_s^\prime)
\hat\Xi(x_s^\prime,z_s^\prime)\rangle
\exp(ik_xx_s^\prime)
\frac{\sin(k_zz_s^\prime)}{k_z}.
\label{M}
\end{eqnarray}

By equating the denominator of Eq.~(\ref{G-Fourier}) to zero, we obtain
the dispersion equation for the rough-bounded strip. The solution to this
equation in the lowest (linear) order in $M(k_x)$ is $k_x=k_n+\delta k_n$.
Here $k_n=\sqrt{k^2-(\pi n/d)^2}$ is the unperturbed longitudinal
wavenumber of an $n$-th propagating normal waveguide mode and
$\delta k_n$ being the complex modification to $k_n$ caused by the wave
scattering from the lower irregular boundary,

\begin{equation}
\delta k_n = \gamma_n + i(2L_n)^{-1}=
-M(k_n)(\pi n/d)^2/k_nd.
\label{dk-M}
\end{equation}
The real part $\gamma_n$ of $\delta k_n$ is responsible for variation of
the phase velocity (for dephasing), while $L_n$ has the meaning of the
attenuation length for the $n$-th mode.

In fact, a form of the solution (\ref{G-Fourier}), (\ref{dk-M}) is common
and well-known~\cite{BF,McGurn}. Our improvement lies in the self-energy
$M(k_x)$ (\ref{M}). This expression is nothing else but a first
non-vanishing (quadratic) term in an expansion of $M(k_x)$ in powers of
the {\it exact} scattering operator (\ref{Xi}). We stress that such an
approximation is essentially different from and is much more general than
an extensively exploited first term in an expansion of
$M(k_x)$ in powers of the dispersion $\zeta^2$.

To find the domain of validity for Eqs.~(\ref{M}), (\ref{dk-M}) one can
use ideas proposed in the book~\cite{Rytov}. We have proved that this
domain coincides with the two natural requirements of {\it weak wave
scattering} from a rough boundary, which are equations themselves with
respect to the external dimensionless parameters $(k\zeta)^2$, $kR_c$,
$kd/\pi$, and $n$,

\begin{equation}
|\delta k_n|\Lambda_n \ll 1, \qquad
|\delta k_n|\widetilde R_c \ll 1.
\label{WS}
\end{equation}
The first of these implies smallness of the complex phase shift over the
distance $\Lambda_n=2k_nd/(\pi n/d)$ passed by an $n$-th mode between two
successive reflections from the rough boundary. Under this condition the
mode experiences a large number $L_n/\Lambda_n \gg 1$ of the reflections
before its amplitude will be substantially attenuated. Also this ensures
smallness of $\delta k_n$ in comparison with the unperturbed wavenumber
$k_n$, because the inequality $k_n\Lambda_n \gtrsim 1$ always holds.

The second of Eqs.~(\ref{WS}) indicates that the phase shift must remain
small over the typical variation scale $\widetilde R_c$ (effective
correlation radius) of the boundary scattering potential. Obviously, this
is the necessary and sufficient condition for correctness of statistical
averaging over $\xi(x)$. The quantity $\widetilde R_c$ is defined as the
typical width of the correlator
$\langle\hat\Xi(x_s,z_s)\hat\Xi(x_s^\prime,z_s^\prime)\rangle$ in
Eq.~(\ref{M}) as a function of $x_s-x_s^\prime$. It does not generally
coincide with the mean length $R_c$ of boundary defects.

We next calculate the correlator and the integrals over $z_s$ and
$z_s^\prime$ in Eq.~(\ref{M}). Then we substitute the result into
Eq.~(\ref{dk-M}) and extract real and imaginary parts of $\delta k_n$.
Finally, we get explicit formulae for $\gamma_n$ and $L_n$:

\begin{eqnarray}
&&\gamma_n = \frac{\zeta^2}{2}\frac{(\pi n/d)^2}{k_n d}
\sum_{n^\prime=1}^{n_d}\frac{(\pi n^\prime/d)^2}{k_{n^\prime} d}
\left[\widetilde W_S(k_n,k_{n^\prime})-
\widetilde W_S(k_n,-k_{n^\prime})\right] -
\frac{(\pi n/d)^2}{k_n d}M_2(k_n) \nonumber \\
&&+2\zeta^2\frac{(\pi n/d)^2}{k_n d}
\sum_{n^\prime=n_d+1}^\infty
\frac{(\pi n^\prime/d)^2}{|k_{n^\prime}| d}
\int_0^\infty dx\exp(-|k_{n^\prime}|x)
{\rm Re}\left[\exp(-ik_n x)
\widetilde{\cal W}(k_n,i|k_{n^\prime}|;x)\right],
\label{gamma}
\end{eqnarray}

\begin{equation}
L_n^{-1} = \zeta^2\frac{(\pi n/d)^2}{k_nd}
\sum_{n^\prime=1}^{n_d}\frac{(\pi n^\prime/d)^2}{k_{n^\prime}d}
\left[\widetilde W_C(k_n,k_{n^\prime})+
\widetilde W_C(k_n,-k_{n^\prime})\right].
\label{L}
\end{equation}
Here integer $n_d=[kd/\pi]$ is the number of the propagating normal modes
in the smooth strip. The function $\widetilde{\cal W}(k_x,q_x;x)$ is the
generalized coefficient of correlation
($\widetilde{\cal W}(k_x,q_x;x) \simeq {\cal W}(|x|)$ as
$\zeta^2 \rightarrow 0$),

\begin{eqnarray}
&&\widetilde{\cal W}(k_x,q_x;x)=
\left(4k_z q_z \zeta^2\right)^{-1}
\nonumber \\
&&\times\Biggl\{
\left[(k_z+q_z)^2+(k_z+q_z)(k_x-q_x)
\left(\frac{k_x}{k_z}-\frac{q_x}{q_z}\right)-
(k_x-q_x)^2 \frac{k_xq_x}{k_zq_z}\right]
S(k_z+q_z,k_z+q_z;x) \nonumber \\
&&-\left[(k_z-q_z)^2+(k_z-q_z)(k_x-q_x)
\left(\frac{k_x}{k_z}+\frac{q_x}{q_z}\right)+
(k_x-q_x)^2 \frac{k_xq_x}{k_zq_z}\right]
S(k_z-q_z,k_z-q_z;x) \nonumber \\
&&+2(k_x-q_x)\left[q_x\frac{k_z}{q_z}-k_x\frac{q_z}{k_z}+
(k_x-q_x)\frac{k_xq_x}{k_zq_z}\right] S(k_z+q_z,k_z-q_z;x)\Biggr\};
\label{Wt-def}
\end{eqnarray}

\begin{equation}
S(t_1,t_2;x) = (t_1t_2)^{-1}
\sinh\left[t_1t_2\zeta^2{\cal W}(|x|)\right]
\exp\left[-(t_1^2+t_2^2)\zeta^2/2\right],
\label{S}
\end{equation}
where $q_z=k_z(q_x)$ and the functions $\widetilde W_{S(C)}(k_x,q_x)$ stand
for sine and cosine Fourier transforms of
$\widetilde{\cal W}(k_x,q_x;x)$ respectively. The component $M_2(k_x)$ of
the self-energy is given by

\begin{eqnarray}
&&M_2(k_x) = \frac{k^2}{2k_z^2d}\sum_{n^\prime=1}^\infty
\Bigl\{2S(k_z+\pi n^\prime/d,k_z-\pi n^\prime/d;0)
\nonumber \\
&&-S(k_z+\pi n^\prime/d,k_z+\pi n^\prime/d;0)-
S(k_z-\pi n^\prime/d,k_z-\pi n^\prime/d;0)
\nonumber \\
&&-\frac{k_z}{k^2}\frac{\pi n^\prime}{d}
\left[S(k_z+\pi n^\prime/d,k_z+\pi n^\prime/d;0)-
S(k_z-\pi n^\prime/d,k_z-\pi n^\prime/d;0)\right]\Bigr\}.
\label{M2}
\end{eqnarray}

An essential distinction between $\gamma_n$ (\ref{gamma}) and $L_n^{-1}$
(\ref{L}) is that the latter is formed by scattering of a given $n$-th
propagating mode into propagating waveguide modes with $n^\prime \leq n_d$
only, while the former has much more complicated structure due to
contributions of both propagating and evanescent ($n^\prime > n_d$) modes.
This feature is a basis for surprising properties of $\gamma_n$.

{\it Brief analysis and discussions}: We start with a relatively simple
and widely used limiting case of {\it small boundary perturbations}, when

\begin{equation}
(k\zeta)^2 \ll 1.
\label{SBP}
\end{equation}
Here Eq.~(\ref{L}) for $L_n$ is simply reduced to the standard result from
the earlier works~\cite{BFF,BF} by replacing
$\widetilde{\cal W}(k_x,q_x;x)$ with ${\cal W}(|x|)$. On the contrary,
$\gamma_n$ (\ref{gamma}) shows an unconventional type of behavior. The
reason is that the last two terms of Eq.~(\ref{gamma}) are mainly formed
by those evanescent modes whose normal wavelengths $(\pi n^\prime/d)^{-1}$
are of the order of the roughness height $\zeta$.  Each of such `resonant'
modes gives a contribution to $\gamma_n$ proportional to $\zeta^2$, while
the number $n^\prime$ of those modes is $\sim d/\zeta \gg n_d$, i.e. it is
inversely proportional to $\zeta$. All this gives a {\it linear}
dependence of $\gamma_n$ (\ref{gamma}) on the roughness height $\zeta$,

\begin{equation}
\gamma_n \sim \zeta \; (\pi n/d)^2/k_nd.
\label{gamma-SBP}
\end{equation}
This formula is the main result of the Letter. It leads to the following
significant conclusions:

1. Since $L_n^{-1} \propto \zeta^2$ as $\zeta^2 \rightarrow 0$, then
$\gamma_n \gg L_n^{-1}$ and, hence, the entire spectrum shift $\delta k_n$
(\ref{dk-M}) turns out to be nearly real, i.e.
$\delta k_n \simeq \gamma_n \propto \zeta$. This means that a signal
propagating through a nearly smooth waveguide is dephased (chaotized) much
earlier (over much shorter distances) than its initial amplitude is damped.

2. From Eq.~(\ref{gamma-SBP}) and item~1 it follows that $\delta k_n$ is a
{\it non-analytic} (square-root) function of the dispersion $\zeta^2$, or
of the binary correlator (\ref{Gauss}):
$\delta k_n \propto (\zeta^2)^{1/2}$.

3. Usually it is believed that the condition (\ref{SBP}) is sufficient to
infer that any long-wave normal mode with $(k_z\zeta)^2 \ll 1$ (i.e. with
$(\pi n/d)^{-1} \gg \zeta$) is mainly scattered into the long-wave modes
as well, $(\pi n^\prime/d)^{-1} \gg \zeta$. This assumption allows
immediate replacement of the exact Dirichlet boundary condition,
formulated on a randomly rough boundary, by an approximate impedance one
laid down on the averaged (deterministic) boundary $z=0$ with the random
impedance $\xi(x)$. We have proved that such a reduction is groundless,
because the `resonant' evanescent modes with $(\pi n^\prime/d)^{-1} \sim
\zeta$ dominate. Thus, the problem of wave propagation through a waveguide
with an absolutely soft random boundary can not be reduced to that with
the smooth random-impedance boundary even for the arbitrarily weak
perturbations.

An important step in analyzing the case (\ref{SBP}) is to obtain the
explicit weak-scattering condition. To this end, we substitute
Eq.~(\ref{gamma-SBP}) into Eqs.~(\ref{WS}) and apply the asymptotic
$\widetilde R_c \simeq R_c$, which is valid for $(k\zeta)^2 \ll 1$. As a
result, we find that Eqs.~(\ref{WS}) can be rewritten as

\begin{equation}
(k_z\zeta)^2 = (\pi n/d)^2\zeta^2 \ll
\min\left\{1,(\Lambda_n/R_c)^2\right\}.
\label{WS-SBP}
\end{equation}
This inequality is automatically satisfied within Eq.~(\ref{SBP}) if
successive reflections of an $n$-th mode from the rough boundary are
{\it not correlated}~\cite{BFF,BF} ($R_c \ll \Lambda_n$). However, if the
correlations are {\it strong} ($\Lambda_n \ll R_c$), then
Eq.~(\ref{WS-SBP}) supplements Eq.~(\ref{SBP}) and may even become
more restrictive than Eq.~(\ref{SBP}). Note that the roughness slope
$\zeta/R_c$ may far exceed unity within the limit~(\ref{SBP}),
(\ref{WS-SBP}).

The situation with {\it large boundary defects}, when

\begin{equation}
(k\zeta)^2 \gg 1,
\label{LBD}
\end{equation}
is much more diverse and complicated than the case (\ref{SBP}). For the
most part it is analyzable only numerically. Therefore we postpone the
details until the longer paper~\cite{MM} and list here only few of
intriguing results:

1. In contrast to the case (\ref{SBP}), the imaginary part of $\delta k_n$
(\ref{dk-M}) may well compete with its real part. Moreover, the situation
with $L_n^{-1} \gtrsim |\gamma_n|$ is rather typical.

2. The real spectrum shift $\gamma_n$ (\ref{gamma}) reverses the sign as
$k\zeta$ reaches some threshold value $k\zeta \approx 1.5 \div 2.5$, which
is slightly dependent on the other parameters. So, the large boundary
defects (\ref{LBD}) may not only decrease, but also increase the phase
velocity of a propagating wave.

3. As $k\zeta \lesssim 2$, $\delta k_n$ is almost insensitive to the slope
$\zeta/R_c$ and to the presence of the shadowing effect, which is
controlled by the Fresnel parameter $k\zeta^2/R_c$. However, as
$k\zeta \gtrsim 2$, the weak-scattering conditions (\ref{WS}) are
fulfilled for not too steep slopes only ($\zeta/R_c \lesssim 2 \div 3$),
while the strong shadowing effect is still allowed
($k\zeta^2/R_c \lesssim 8 \div 10$).

We mention that the approach presented in this Letter can be extended to
vector wave fields.

\vspace{12pt}

The authors thank Prof.~A.~A.~Maradudin and Dr I.~V.~Yurkevich for useful
comments on this work.

\end{document}